\documentclass[twocolumn]{aastex62}
\usepackage{graphicx}                    
\usepackage{times}                    
 \bibpunct{(}{)}{;}{a}{}{,} 
\usepackage{amssymb}                    
\usepackage{color}                       
\usepackage{url}                         
\usepackage{amsmath} 

\def\beg{\begin{eqnarray}}
\def\ende{\end{eqnarray}}
\def\lsim{\lower.4ex\hbox{$\;\buildrel <\over{\scriptstyle\sim}\;$}}
\def\gsim{\lower.4ex\hbox{$\;\buildrel >\over{\scriptstyle\sim}\;$}}

\newcommand{\Pm}{\mbox{Pm}}
\newcommand{\Prr}{\mbox {Pr}}
\newcommand{\Rm}{\mbox{Rm}}

\newcommand{\Mm}{\mbox{Mm}}
\renewcommand{\vec}[1]{\mbox{\boldmath $#1$}}

\def \n   {\vec{\nabla}\!}

\def \Om  {{\it \Omega}}

\def \etaT{\eta_ {\rm T}}
\def \kappaP{\kappa_ {\rm P}}

\def\A{Alfv\'en}

\def\gsim{\lower.4ex\hbox{$\;\buildrel >\over{\scriptstyle\sim}\;$}} 
\def\lsim{\lower.4ex\hbox{$\;\buildrel <\over{\scriptstyle\sim}\;$}} 

\renewcommand{\vec}[1]{\mbox{\boldmath $#1$}}

\def\aap{Astronomy \& Astrophysics}

\def\ara\&a{Ann. Rev. Astronomy Astrophysics}

\def\apjs{Astrophys. J. Suppl.}

\begin{document}



\title{The turbulent pressure of  magnetoconvection  for slow and rapid rotation}
\shorttitle{Pressure of rotating magnetoconvection}
%
\correspondingauthor{M. K\"uker}
\email{mkueker@aip.de}

\author{M.~K\"uker} 

\author{G. R\"udiger}

   \affiliation{Leibniz-Institut f\"ur Astrophysik Potsdam (AIP), An der Sternwarte 16, 14482 Potsdam, Germany}     
   
 \shortauthors{M. K\"uker and G. R\"udiger}  


%

 
\begin{abstract}
Motivated by recent simulations of  sunspot formation, we extend the theory of  the pressure difference between magnetized and non-magnetized gas by Dicke to include rotating turbulence.   While the (vertical)  background field  provides a  positive-definite magnetic pressure difference between the magnetized and the unmagnetized gas, Reynolds stress and Maxwell stress of turbulence strongly modify this result.  With  the quasilinear approximation we demonstrate that the influence of the turbulence differs between the high-conductivity and the low-conductivity limits. 
Sufficiently small magnetic Reynolds numbers  lead  to magnetic pressure suppression where indeed the pressure  excess can even  assume negative values. Box simulations of magnetoconvection subject to a vertical magnetic field carried out with the Nirvana code confirm this overall picture. They also demonstrate how a global rotation  {\em reduces}  the negative magnetic pressure effect. For rapid rotation  the  total magnetic pressure difference  caused by large-scale magnetic fields { and}  turbulence even fully disappears  for small field strengths. Magnetic  fields of moderate strength thus neither reduce nor enhance the  turbulence pressure  of rapidly  rotating convection. Consequences of this phenomenon for the star formation efficiency  are shortly discussed. 
\end{abstract}

\keywords{Magnetohydrodynamics (MHD) --  Magnetic fields  -- rotation --   convection -- sunspots}

%
\section{Introduction} \label{Section1}
If in a fluid a magnetized domain  exists with a non-magnetic surrounding in  quasistationary equilibrium and the magnetic field has   a cylindrical geometry $\vec{B}=(0,0, B_0(R,\phi))$ then the gas pressure and the temperature differences between the inner and the outer regions will simply be 
\beg
\delta P= -\frac{B_0^2}{2\mu_0}, \ \ \ \ \ \ \ \ \ \ \ \ \ \ \ \ \ \ \delta T= -\frac{B_0^2}{2\mu_0}\frac{T}{P},
\label{Dicke1}
\ende
\citep{D70}. The magnetic domain is thus cooler than its non-magnetic surrounding.
This straightforward explanation of the sunspot phenomenon fails, however, in one basic  respect. Both the magnetized domain {\em and} its non-magnetic surrounding are turbulent. The large-scale  Maxwell stress -- which was the only one considered by Dicke -- has thus to be completed by the Reynolds stress and the small-scale Maxwell stress caused by the overall convection. Instead of (\ref{Dicke1}) we write
\beg
P_{\rm in}=P_{\rm ext} -\delta P^{\rm tot},
\label{Dicke2}
\ende
where $P_{\rm in}$ is the pressure in the magnetic region, $P_{\rm ext}$ the pressure outside the magnetic region, and 
the second term on the right hand side collects the contributions from Reynolds stress and Maxwell stress. Note that the sign of $\delta P^{\rm tot}$ is not known from the definition. Positive values cause the gas pressure of the magnetized fluid to be reduced compared with  the external pressure as described by Eq. (\ref{Dicke1}). If, however,  $\delta P^{\rm tot}<0$ would occur  because of the influence of turbulence, then  the inner gas pressure would exceed the external gas pressure, with consequences recently  described by \cite{LW17}  with respect to the theory of sunspot formation. Our computations confirm the existence of this phenomenon, provided the magnetic field is not too strong, the rotation not too rapid, and the electric conductivity not too high. We find that rapid rotation  leads to  $\delta P^{\rm tot}\simeq  0$, so that the   inner gas pressure approaches the outer gas pressure, which according to the second relation in (\ref{Dicke1}) will  also be  true for the temperature. 

In the following we derive both the turbulence-originated Reynolds stresses and the Maxwell stresses for a turbulent fluid rotating with an angular velocity $\vec{\Om}$ under the presence of a uniform background field $\vec B$. The magnetic field will here be considered as vertical, i.e. as antiparallel to the density gradient. The fluctuating flow   and   field components  are denoted by $\vec{u}$ and  by $\vec{b}$, resp. The  standard Maxwell stress tensor
\beg
  M_{ij}= \frac{1}{\mu_0} B_iB_j - \frac{1}{2\mu_0}\vec{B}^2 \delta_{ij}
  \label{2}
\ende
of the considered MHD turbulence turns into the generalized stress tensor
\beg
  M_{ij}^{\rm tot}= M_{ij} -\rho Q_{ij} +M_{ij}^{\rm T},
  \label{3}
\ende
with the one-point correlation tensor 
\beg
Q_{ij}= \langle u_i(\vec{x},t) u_j(\vec {x},t)\rangle
 \label{33}
\ende
of the flow
and the turbulence-induced  Maxwell stress tensor
\beg
 M_{ij}^{\rm T}= \frac{1}{\mu_0} \langle b_i(\vec{x},t)b_j(\vec{x},t)\rangle - \frac{1}{2 \mu_0}
\langle \vec{b}^2(\vec{x},t)\rangle \delta_{ij}.
  \label{3b}
\ende
The terms with the Kronecker deltas  in (\ref{3}) form the total pressure. We write the one-point correlation tensor (\ref{33}) for homogeneous and isotropic turbulence as
\beg
Q_{ij}=q_1\delta_{ij} + q_2 \Om_i\Om_j +q_3 B_iB_j.
 \label{qs}
\ende
Each part of this tensor must be even in $\vec\Om$ and even in $\vec B$, there are no mixed elements. Only $q_1$ is thus  relevant for the pressure evaluation. Similarly, for the Maxwell stress tensor, i.e.
\beg
B_{ij}=Q_1\delta_{ij} + Q_2 \Om_i\Om_j +Q_3 B_iB_j ,
 \label{Qs}
\ende
so that for the pressure
\beg
\frac{P^{\rm tot}}{\rho}=q_1 - \frac{Q_1}{\mu_0\rho} +  \frac{B_0^2}{2\mu_0\rho}  +  \frac{1}{2\mu_0\rho}(\langle b_x^2+ b_y^2+b_z^2 \rangle)
 \label{p1}
\ende
results. 
The orientation of the vectors $\vec{\Om}$ and $ \vec{B}$ defines the model. A {\em vertical} magnetic background field is considered which shows in the positive radial ($z$) direction\footnote{$x$ shows in  meridional direction, $y$ in azimuthal direction}, $\vec{B}=B_0\ (0,0,1)$.The angle $\theta$ between the rotation vector and the gravity defines the components of the rotation vector $\vec{\Om}= \Om\  (-\sin\theta,0,\cos\theta)$.
From (\ref{qs}) one obtains 
\begin{equation}
\langle u_x^2  \rangle= q_1 + q_2 \Om^2 \sin^2\theta, \ \ \ \langle u_y^2  \rangle= q_1 + q_2 \Om^2 \cos^2\theta,
 \label{qqs}
\end{equation}
and from (\ref{Qs})
\begin{equation}
\langle b_x^2  \rangle= Q_1 + Q_2 \Om^2 \sin^2\theta, \ \ \ \langle b_y^2  \rangle= Q_1 + Q_2 \Om^2 \cos^2\theta .
 \label{QQs}
\end{equation}
Hence
\begin{eqnarray}
\frac{P^{\rm tot}}{\rho} &=&
\frac{\cos^2\theta\langle u_x^2\rangle- \sin^2\theta\langle u_y^2\rangle}{\cos^2\theta- \sin^2\theta} \nonumber \\
&&+\frac{B_0^2}{2\mu_0\rho} 
+ \frac{\langle b_x^2+b_y^2+b_z^2\rangle}{2\mu_0\rho} \nonumber \\
 &&-\frac{1}{\mu_0\rho}\frac{\cos^2\theta\langle b_x^2\rangle- \sin^2\theta\langle b_y^2\rangle}{\cos^2\theta- \sin^2\theta}.
  \label{p2}
\end{eqnarray}
Without rotation this expression simplifies to 
\beg
\frac{P^{\rm tot}}{\rho}= \langle u_x^2\rangle
 + \frac{1}{2\mu_0\rho} B_0^2
 + \frac{1}{2\mu_0\rho} \langle b_y^2+b_z^2-b_x^2\rangle
\label{ptot}
\ende
\citep{RK13}. Because of the isotropy in the horizontal plane we have $\langle b_y^2+b_z^2-b_x^2\rangle = \langle b_z^2\rangle$. All  terms on the RHS of (\ref{ptot}) are thus  positive.
 For large magnetic background fields the middle term will  exceed the other terms and for large magnetic Reynolds numbers the last term will dominate.  Nevertheless, for weak fields and magnetic Reynolds numbers not too large it happens that the magnetic quenching of the first term in (\ref{ptot}) provides so small intensities that the total magnetic-influenced pressure becomes smaller than the  turbulence intensity $\langle u_x^{(0)2}\rangle $ hence for the gas pressure $P_{\rm in}>P_{\rm ext}$ instead of (\ref{Dicke1}) which has been called the negative-pressure phenomenon \citep{BK10,BK11,BK12}. In the simplest geometry after (\ref{Dicke1}) then also $T_{\rm in}>T_{\rm ext}$ holds for the temperature.

With both  analytical and  direct numerical simulations \cite{LB13} calculated the total magnetic pressure (\ref{p1}) for driven turbulence with   weak horizontal magnetic fields  under the presence of global rotation. The  rotation, however, was too  slow   for the numerical simulations  to show  trends for suppression or amplification of the total magnetic-influenced turbulence pressure (their Fig. 1).


\section{Driven  turbulence}\label{driven}
The  influences of the magnetic field and the basic rotation on driven turbulence are similar but not identical. While the magnetic field always reduces the velocity components perpendicular to the magnetic field  this is not obvious for rotation \citep{C61}. In order to demonstrate this situation we shall in the following formulate a quasilinear theory of the modifications a turbulence field  driven by a fluctuating force field undergoes through the combined action of a uniform magnetic field and solid-body rotation. For simplicity the magnetic field may be aligned with the rotation axis. Because of the structure of the pressure equation (\ref{ptot}) we are  particularly interested in results for the transverse intensity $\langle u_x^2\rangle$ when the magnetic field and the rotation axis define the $z$-axis.

It is almost trivial that within  this concept the signs of $\vec{\Om}$ and $\vec{B}$ do not play any role.  The Fourier component of the velocity field in the rotating system under the influence of magnetic fields can  be expressed via 
\beg   
 \hat{u}_i({\vec k},\omega) = D_{ij}\hat{u}^{(0)}_j({\vec k},\omega)
 \label{D}
 \ende 
 by the Fourier component of the driven turbulence $\hat{\vec{u}}^{(0)}({\vec k},\omega)$.
 The tensor $D_{ij}$ for the simultaneous influence of global rotation and magnetic field, 
\beg
    D_{ij} = \frac{N\delta_{ij} + W\epsilon_{ijl}k^\circ_l}
    {N^2 + W^2},
    \label{Dij}
\ende
has been given by \cite{KR94}. Here $N$ carries the impact of the magnetic field and $W$ that of the global  rotation. In detail it  is 
\begin{eqnarray}
    N &=& 1 + \frac{(\vec{ k}\cdot\vec {V}_{\rm A})^2}
    {(-i\omega +\nu k^2)(-i\omega+\eta k^2)},
    \nonumber \\
    W &=& \frac{2({\vec k}^\circ\cdot{\vec\Omega})} {-i\omega +\nu k^2 }. 
    \label{NW}
\end{eqnarray}
As required,   all terms are invariant against the transformation $\vec{B}\to -\vec{B}$,  the  sign of the magnetic field does not play any role. $\nu$ is the microscopic viscosity,   $\eta=1/\mu_0\sigma$ the magnetic resistivity, $\sigma$ the electric conductivity and  $V_{\rm A}=B_0/\sqrt{\mu_0\rho}$   the \A\  velocity.  ${\vec k}^\circ = {\vec k}/k$ is a unit vector. 

For homogeneous and isotropic turbulence with its  spectral tensor, 
\begin{eqnarray}
    \hat{Q}^{(0)}_{ij}({\vec k},\omega ) &=& \frac{E(k,\omega)}{16\pi k^2}
    (\delta_{ij} - k^\circ_i k^\circ_j), 
    \label{4}
\end{eqnarray}
we find after some algebra that 
\beg
 &&\hat{Q}_{ij}({\vec k},\omega ) =
    \frac{E(k,\omega)}{16\pi k^2 (N^2 + W^2)({N^*}^2 + {W^*}^2)} \times     \nonumber \\
    &&   [(N{N^*} + W{W^*})(\delta_{ij} - k^\circ_i k^\circ_j)   \nonumber \\
    &&\hspace{1.5cm}+({N^*}W - N{W^*})\epsilon_{ijl}k^\circ_l] \label{5}
    \ende
with the asterisk for complex conjugate expressions. 
The local spectrum $E$ is defined by
\beg
\langle \vec{u}^2\rangle =\int_0^\infty\int_0^\infty E(k,\omega)\ {\rm d}{k}\ {\rm d}\omega.
\label{ux4}
\ende

For weak fields and slow rotation it is thus enough to consider the correlation tensor in the first order of $B_0^2$ and $\Om^2$.  One finds 
\beg
&&\langle u_x^2\rangle- \langle u_x^{(0)2} \rangle =\iint (k_y^2+k_z^2) \Big(\frac{3\omega^2-\nu^2k^4}{(\omega^2+\nu^2k^4)^2}(2 \vec{k}\cdot\vec{\Om}/k)^2+\nonumber\\
&&\ \ \ 
+  \frac{2(\omega^2-\nu\eta k^4)}{(\omega^2+\nu^2k^4)(\omega^2+\eta^2k^4)} (\vec{k}\cdot\vec{V}_{\rm A})^2 \Big) \times \nonumber \\
 && \hspace{1.5cm}  \frac{E}{16\pi k^4}\ \ {\rm d}\vec{k}\ {\rm d}\omega,
\label{ux2}
\ende
which for $\nu=\eta$ can also be written as  
\beg
\lefteqn{\langle u_x^2\rangle- \langle u_x^{(0)2} \rangle =\frac{1}{16\pi} \iint \frac{k_y^2+k_z^2}{k^4}}\nonumber\\
&&\ \ \  \ \ \ \ \ \ \ \ \  \Big(\frac{\omega}{(\omega^2+\nu^2k^4)^2}\frac{\partial (\omega^2+\nu^2 k^4) E}{\partial \omega}
(2 \vec{k}\cdot\vec{\Om}/k)^2+\nonumber\\
&&\ \ \ \ \  \ \ \ \ \ \ \ \ 
+  \frac{2\omega}{\omega^2+\eta^2k^4}\frac{\partial  E}{\partial \omega} (\vec{k}\cdot\vec{V}_{\rm A})^2 \Big) \ \ {\rm d}\vec{k}\ {\rm d}\omega.
\label{ux30}
\ende 
The coefficients of both terms are different. The influences of rotation and magnetic field onto turbulence are thus not identical.  For flat `white-noise` spectra $E\simeq$~const the (weak) magnetic field does not affect the turbulence intensity while the rotation leads to an  `anti-quenching' of the transverse turbulence intensity $\langle u_x^2 \rangle$. On the other hand, for  spectra $E$ which are steep enough both effects are suppressing (`quenching') the turbulence. This finding does not change for $\nu\neq \eta$, see Eq. (37) of \citet{R74}.

For the flows parallel to the rotation axis the expression
\beg
&&\langle u_z^2\rangle- \langle u_z^{(0)2} \rangle =    \frac{1}{16\pi} \iint \frac{k_x^2+k_y^2}{k^4}\nonumber\\
&&\ \ \  \ \ \ \ \ \ \ \ \  \Big(\frac{\omega}{(\omega^2+\eta^2k^4)^2}\frac{\partial (\omega^2+\eta^2 k^4) E}{\partial \omega}
(2 \vec{k}\cdot\vec{\Om}/k)^2\nonumber\\
&&\ \ \ \ \  \ \ \ \ \ \ \ \ 
+  \frac{2\omega}{\omega^2+\eta^2k^4}\frac{\partial  E}{\partial \omega} (\vec{k}\cdot\vec{V}_{\rm A})^2 \Big) \ \ {\rm d}\vec{k}\ {\rm d}\omega
\label{ux3}
\ende 
results and for the anisotropy caused by rotation and magnetic field
\beg
\lefteqn{\langle u_z^2\rangle- \langle u_x^2 \rangle =\frac{1}{16\pi} \iint 
\frac{k_x^2-k_z^2}{k^4}\frac{\omega}{\omega^2+\eta^2 k^4}}\nonumber\\
&&\ \ \  \ \ \ \ \ \ \ \ \  \Big(\frac{1}{\omega^2+\eta^2k^4}\frac{\partial (\omega^2+\eta^2 k^4) E}{\partial \omega}
(2 \vec{k}\cdot\vec{\Om}/k)^2+\nonumber\\
&&\ \ \ \ \  \ \ \ \ \ \ \ \ 
+  2\frac{\partial  E}{\partial \omega} (\vec{k}\cdot\vec{V}_{\rm A})^2 \Big) \ \ {\rm d}\vec{k}\ {\rm d}\omega,
\label{ux41}
\ende 
always for $\Pm=1$. After integration over the wave number components we find
\beg
&&\langle u_z^2\rangle- \langle u_x^2 \rangle =-\frac{1}{60\pi} \iint 
\frac{\omega}{(\omega^2+\eta^2 k^4)k^2}\nonumber\\
&&   \Big(\frac{2\Om^2}{\omega^2+\eta^2k^4}\frac{\partial (\omega^2+\eta^2 k^4) E}{\partial \omega}
+  k^2V_{\rm A}^2\frac{\partial  E}{\partial \omega}\Big)  {\rm d}\vec{k}\ {\rm d}\omega
\label{ux5}
\ende 
for an isotropic spectral function $E=E(k,\omega)$. Indeed, the rotational and the magnetic influences  are going in different directions.
While the magnetic field supports the vertical turbulence intensity the rotation supports the horizontal intensity. This behavior, which was already suggested by \cite{C61}, becomes finally clear after inspection of the relation
\begin{equation}
\langle u_z^2\rangle- \langle u_x^2 \rangle =\frac{\Om^2}{15\pi}\iint
\frac{k}{\omega^2+\eta^2 k^4}
\frac{\partial }{\partial k}\Big(\frac{E}{k^2}\Big)  \ \ {\rm d}\vec{k}\ {\rm d}\omega,
\label{ux6}
\end{equation}
 equivalent to the rotation-induced part in (\ref{ux5}). As the wave number derivative in (\ref{ux6}) is certainly negative it follows $\langle u_x^2\rangle> \langle u_z^2 \rangle$. This is opposite to the  influence of magnetic fields in (\ref{ux5}) which runs with $-\partial E/\partial \omega>0$ hence reducing the horizontal flow components  compared with the vertical  ones.
On the other hand, it can be shown  with the expression  (\ref{5}),  that for  $\Om\to \infty$ or $V_{\rm A}\to \infty$  always $\langle u_z^2\rangle=2  \langle u_x^2 \rangle$  \citep{R74}.  In both cases, for sufficiently rapid rotation or strong magnetic field, the vertical turbulence intensity  exceeds the 
transverse intensity by a factor of two.

Magnetic fluctuations also contribute to the magnetic pressure (\ref{ptot}). Horizontal isotropy implies that we only have to compute $\langle b_z^2\rangle$. To the first order in $B_0^2$ one finds the simple relation
\beg
\langle b_z^2\rangle =\frac{B_0^2}{120\pi} \iint 
\frac{E}{\omega^2+\eta^2 k^4}
  \ \ {\rm d}\vec{k}\ {\rm d}\omega,
\label{bz1}
\ende 
so that after (\ref{ptot}) the pressure difference between the magnetic and non-magnetic turbulence is 
\begin{equation}
\frac{\delta P^{\rm tot}}{\rho}= \frac{V^2_{\rm A}}{2}\Big(1
 - \frac{1}{30\pi}   \iint 
\frac{7\eta^2 k^4-9\omega^2}{(\omega^2+\eta^2 k^4)^2} E(k,\omega)
  \  {\rm d}\vec{k}\ {\rm d}\omega\Big).
 \label{ptot2}
\end{equation}
The turbulence part in this expression does {\em not} have a definite sign.

  Equation (\ref{ptot2}) may also be written as 
\beg
\frac{\delta P^{\rm tot}}{\rho} =  (1-\kappa_{\rm P})\ \frac{V^2_{\rm A}}{2}, 
 \label{ptot30}
\ende
hence
\beg
P_{\rm in}=  P_{\rm ext}-(1-\kappa_{\rm P})\rho \frac{V^2_{\rm A}}{2}, 
 \label{ptot30a}
\ende
Without turbulence  $\kappaP=0$, see Eq. (\ref{Dicke1}). Positive $\kappaP$  reduces the pressure difference.  If even  $\kappaP>1$ the inner gas pressure  exceeds the outer one \citep{KR89}. On the other hand, anti-quenching of the turbulence by the magnetic field  leads to negative $\kappaP$ which amplifies  the magnetic suppression of the inner gas pressure \citep{RS75}. 

Positive $\kappaP$ are a necessary condition for the occurrence of a negative-pressure effect.   We have  thus mainly to discuss the  sign of $\kappaP$ which basically  depends on the form of the spectral function $E$.
For $\Pm=1$ one finds 
\begin{equation}
\kappaP= \frac{1}{30\pi}   \iint  \frac{7\eta^2 k^4-9\omega^2 }{(\omega^2+\eta^2 k^4)^2} E(k,\omega)  {\rm d}\vec{k}\ {\rm d}\omega
 \label{ptot31a}
\end{equation}
or, what is the same, 
\begin{equation}
\kappaP= 
 - \frac{1}{30\pi}   \iint 
(\frac{7\omega}{\omega^2+\eta^2 k^4}\frac{\partial E}{\partial \omega} + \frac{2\omega^2 E}{(\omega^2+\eta^2 k^4)^2} )
   {\rm d}\vec{k}\ {\rm d}\omega.
 \label{ptot31}
\end{equation}
Negative values of $\delta P^{\rm tot}$  require  large positive values of $\kappa_{\rm P}$ which is only possible for very steep  $E(k,\omega)$ as a function of $\omega$. 
For a steep function such as  $\delta(\omega)$  the integral is positive and can even lead to negative  $\delta P^{\rm tot}$ for small magnetic Reynolds numbers 
\beg
\Rm' = \frac{u_{\rm rms} \ell_{\rm corr}}{\eta}
\label{Rm}
\ende
 of the turbulence (`low-conductivity limit').

The two terms in  Eq. (\ref{ptot31}) are easy to understand. The first one certainly vanishes for `white noise' ($E\simeq $~const)  and the second one comes from the positive-definite contribution of the Maxwell stress (\ref{bz1}). 
For the white-noise spectrum  the total pressure  excess (\ref{ptot30}) {\em cannot}  become negative.  
 Similarly, we expect  the integral in (\ref{ptot31a})  to be negative  for  small values of $\eta$. In the high-conductivity limit, $\eta\to 0$,
\beg
   \int_0^\infty 
\frac{7\eta^2 k^4-9\omega^2}{(\omega^2+\eta^2 k^4)^2} E(k,\omega)
 \ {\rm d}\omega =-\frac{\pi}{2\eta k^2} E(k,0),
 \label{ptot32}
\ende
hence the leading term  of $\kappaP$ is negative\footnote{$E(k,0)$ vanishes only for undamped waves}   so that  always ${\delta P^{\rm tot}}/{\rho}> {V^2_{\rm A}}/{2} $. 

Hence, in the quasilinear approximation a negative turbulent pressure excess   can only exist in the low-conductivity limit   if $\eta$ is not too small. 
The value of $\kappaP$ is negative in the high-conductivity limit  and it is positive  in the low-conductivity limit. 
Only in the latter case   the negative-pressure effect can  appear. More exactly speaking, the total turbulent pressure excess can only be  negative for
 $
 \Rm'
 <\pi^2 w^*_0
$
 if, in the sense of the mixing length theory, $\tau_{\rm corr}\simeq \ell_{\rm corr}/u_{\rm rms}$.
We have shown earlier with  simplified  spectral functions that  the characteristic value of $w_0^*$  is of the order of 10 for $\Pm\simeq 1$  \citep{RKS12}. 
For large magnetic Reynolds numbers $\Rm'$ of the fluctuations, the integral (\ref{ptot31a}) becomes negative and the effective pressure is enhanced rather than reduced. The bottom plot of Fig. 8 in \cite{KB12} demonstrates  the decrease of positive $\kappaP$  of increasing $\Rm'$ (for forced turbulence).

It remains to consider  $\Pm\neq 1$ for which Eq. (\ref{ptot31a}) turns into
\beg
\kappaP  = \frac{1}{30\pi}   \iint
 \frac{((8\eta-\nu)\nu k^4-9\omega^2) E}{(\omega^2+\nu^2 k^4)(\omega^2+\eta^2 k^4)} 
   {\rm d}\vec{k}\ {\rm d}\omega.
 \label{ptot34}
\ende
It  is clearly negative-definite for $\Pm\geq 8$. For large magnetic Prandtl number, therefore, the pressure { grows} by the turbulence. For all $\Pm$ it also grows for flat spectra  such as white noise as i
\beg
   \int_0^\infty 
\frac{(8\eta-\nu)\nu k^4-9\omega^2}{(\omega^2+\nu^2 k^4)(\omega^2+\eta^2 k^4)} 
 \ {\rm d}\omega =-\frac{\pi}{2\eta k^2}
 \label{ptot35}
\ende
is negative-definite. The flat parts of the spectrum thus always {\em increase}  the magnetic-induced pressure. SOCA provides negative pressure excesses only for $\Pm<8$ and for  low enough electric conductivities.
After (\ref{ptot34})  $\kappaP$ is positive if the frequency spectrum contains a delta function $\delta(\omega)$ (and $\Pm<8$) and it is always negative for white noise. For a certain spectrum between the considered two extremes the $\kappaP$ will change its  sign.

 \section{Rotating magnetoconvection}
One of the   results of the foregoing Section is the strong dependence of the integrals  for $P^{\rm tot}$ on the form of the spectral function of the turbulence  or, what is the same, on the ratio of the diffusion time scales.  Numerical simulations are needed for further insights. We thus perform simulations for magnetoconvection for various amplitudes of the magnetic field, the ordinary Prandtl number $\Prr=\nu/\chi$ (with  $\chi$ the thermal diffusion coefficient) and  the magnetic Prandtl number $\Pm=\nu/\eta$.  The field $B_0$ is assumed as vertical and homogeneous. 
The  simulations are done with the {\sc Nirvana} code by \cite{Z02}, which uses a conservative finite difference scheme in Cartesian coordinates. The length scale is defined by the depth of the convectively unstable layer.  We assume an ideal, fully ionized gas  heated from below and keep  a fixed temperature at the top of  box. Periodic boundary conditions are formulated in the horizontal plane. The upper and lower boundaries are impenetrable and stress-free. 
The initial state is convectively unstable in the upper half of the box.  Convection sets in if the Rayleigh number
 exceeds its critical value. The model complies with  that of \cite{KBK12} with the main difference of the orientation of the mean magnetic field. As in \cite{KK10} and \cite{KB16} our field is vertically directed  while \cite{KBK12} work with  horizontal mean magnetic fields.
 
\begin{figure}
 \centering
\includegraphics[width=6cm]{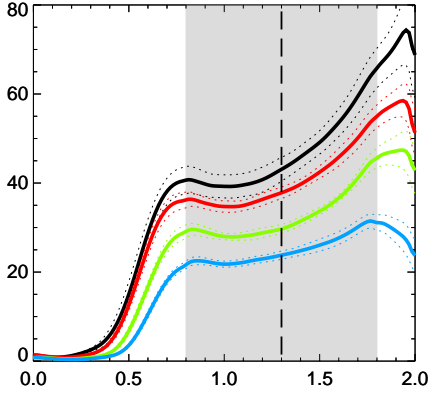}
\caption{The influence of the  molecular Prandtl number $\Prr$ on the turbulent pressure (\ref{ptot}) in units of $(c_{\rm ac}/100)^2$  without rotation and magnetic field.  $\Prr=0.1$ (blue), $\Prr=0.05$ (green), $\Prr=0.02$ (red) and $\Prr=0.01$ (black). The dashed  vertical line marks the center  of the unstable domain in $z$ where 
the blue curve yields  a minimum value of  $\Rm'\simeq 40$, the values for  the other models  are slightly higher. 
$\Om^*=B^*=0$.}
\label{fig11}
\end{figure} 
\begin{figure}
 \centering
\includegraphics[width=6cm]{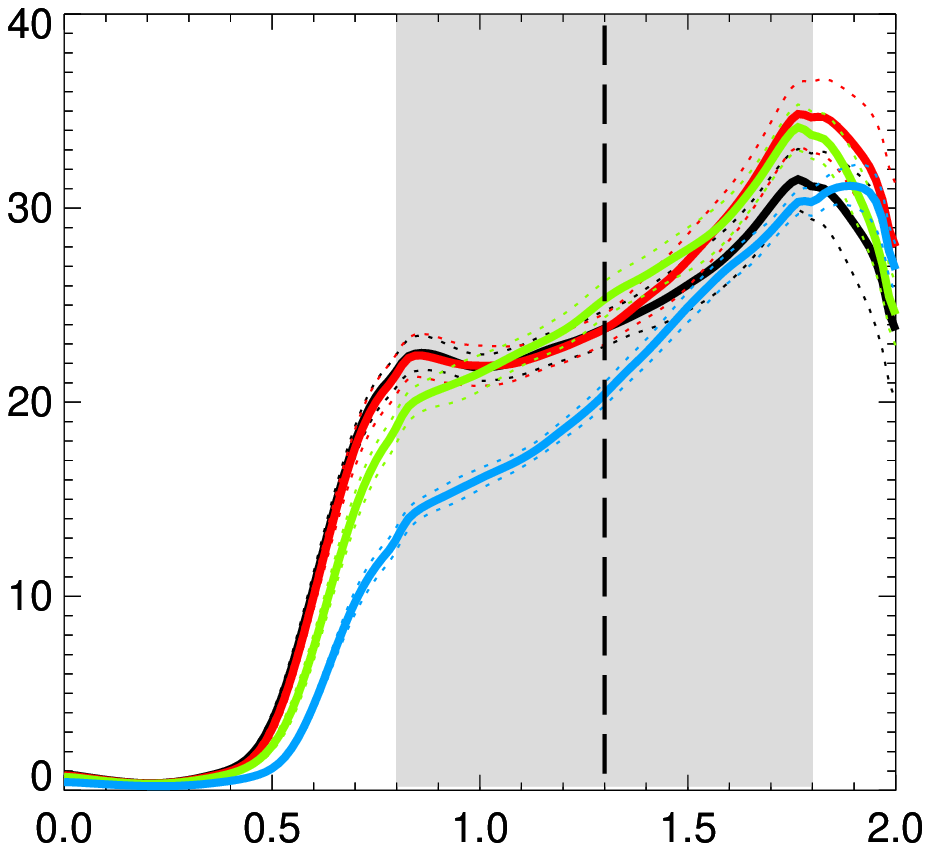}
\includegraphics[width=6cm]{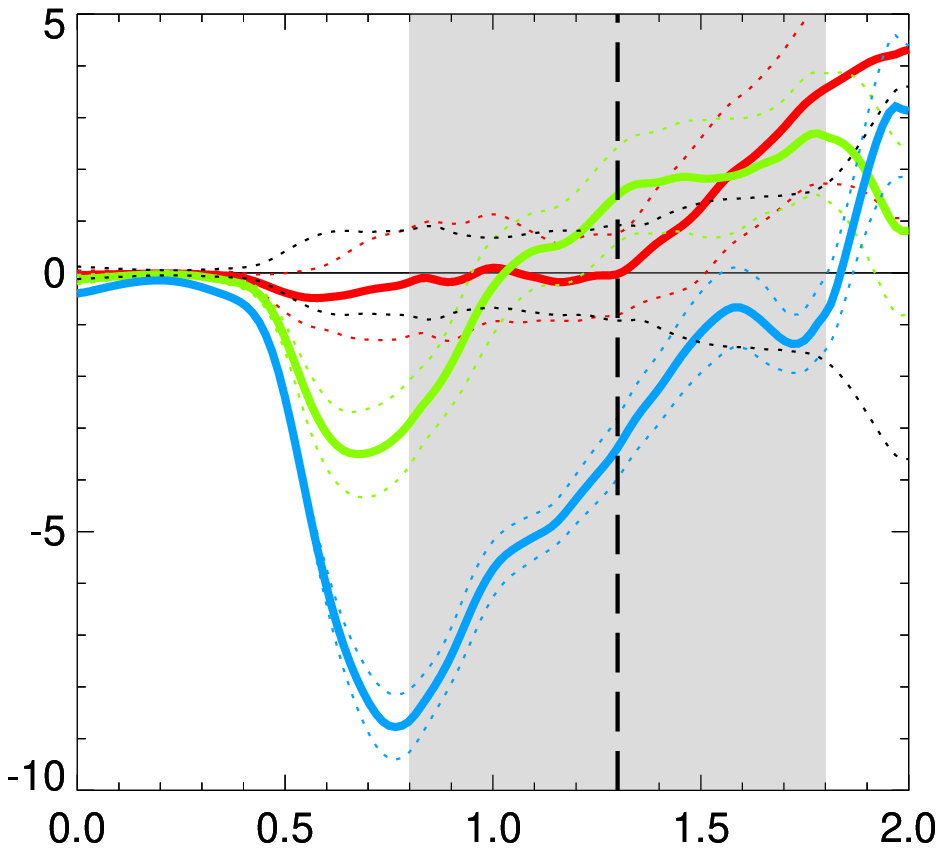}
\caption{The turbulent pressure $\langle u_x^2\rangle$ (top) and its excess  $\delta P^{\rm tot}/\rho= \langle u_x^2\rangle- \langle u_x^{(0)2}\rangle$ (bottom) of  rotating convection in units of $(c_{\rm ac}/100)^2$. Only the blue line is the result of rotational quenching while the red and the green lines stand for rotational anti-quenching. Pressure is rotationally anti-quenched for slow rotation and it is quenched only for rapid rotation. $\Om^*=0$ (black), $\Om^*=1$ (red), $\Om^*=3$ (green) and $\Om^*=10$ (blue). 
$B^*=0$, $\Prr=0.1$.  }
\label{fig10}
\end{figure} 

The velocity field is measured in units of $c_{\rm ac}/100$. This quantity is also used to define the magnetic field via $B_z=B^* \sqrt{\mu_0\rho_0}c_{\rm ac}/100$. In this normalization $B^*=1$ represents 1 kG if   $c_{\rm ac}=10$~km/s and $\rho_0\simeq 10^{-2}$~g/cm$^3$ are adopted. With the equipartition field $B_{\rm eq}=\sqrt{\mu_0\rho_0}u_{\rm rms}$ it is
\beg
\frac{B_z}{B_{\rm eq}}=\frac{B^* c_{\rm ac}}{100 u_{\rm rms}},
 \label{Beq}
\ende
so that for  $u_{\rm rms}=c_{\rm ac}$ the parameter $B^*=10$ describes the moderate  field strength  $B_z=0.1 B_{\rm eq}$. $B^*=10$ for subsonic turbulence with  $u_{\rm rms}=0.1 c_{\rm ac}$ stands for equipartition, $B_z= B_{\rm eq}$. We shall often use in what  follows $B^*=10$
as the value of the prescribed background field. 

By the units of the vertical size $D$ of the convection box and the convective velocity the rotation is normalized by the relation $\Om=\Om^* c_{\rm ac}/100 D$.
If the correlation time of the turbulence is written as $\tau_{\rm corr}={\hat \tau}  D/(c_{\rm ac}/100)$  then 
  $\Om^*\simeq \Om \tau_{\rm corr}/{\hat\tau}$. We shall demonstrate with  simulations that $\hat\tau\simeq 0.1-0.2$.  With the solar values $c_{\rm ac}\simeq 10$~km/s and $D\simeq 200.000$~km the result is  $\Om^*\simeq 4$. 
  
The combination of the rotation rate and the magnetic field yields the magnetic Mach number 
\beg
{\Mm}=\frac{\Om^*}{B^*}
\label{Mm} 
\ende
as a normalized rotation rate which in astrophysical applications often  exceeds unity.  Galaxies have $\Mm\lsim 10$, for the solar tachocline with a magnetic field of 1 kG one obtains $\Mm\simeq 30$, and for typical white dwarfs and neutron stars $\Mm\simeq 1000$ (except magnetars). 

Also the  cores of cold molecular clouds have large magnetic Mach numbers. Polarization measurements of magnetic 
fields in cloud cores  indicate that 
the cores are filled by magnetic fields  of a few $\mu$G \citep{LD09}. 
Velocity measurements  indicate rotation 
rates $\Om\simeq 10^{-13}$s$^{-1}$   for   cloud cores with a radius of 0.1 pc  \citep{BB00,KW09}. 
 With the typical density $10^5$cm$^{-3}$ one finds $\Mm\simeq 5$.

On the other hand,  for stellar material the heat-conductivity $\chi$ is the dominant diffusivity \citep{HG16}.    It is thus  the Roberts number 
$
 { \rm q}={\chi}/{\eta}={\Pm}/{\Prr}\gg1.
 $
 Large $\rm q$ mimic large electric conductivities   if the heat-conductivity  is fixed. Characteristic values  are $\Pm\simeq 6\cdot 10^{-2}$ and $\Prr\simeq 2\cdot 10^{-6}$ so that $\rm q= O(10^4)$ for the bottom of the solar convection zone (see \cite{O03,G07}).  Close to the solar surface the Roberts number becomes smaller but remains  larger than unity. 
\subsection{Rotating convection}
We start to compute the reference pressure of the model without rotation and magnetic field after (\ref{ptot}). By use of the same numerical model \cite{RK12}  in their Fig. 6 found for  weak magnetic field that $u_{\rm rms}\simeq 9$ for the normalized turbulence intensity leading to pressure values of $\simeq 27$. This value is perfectly  fitted by the data given in  Fig. \ref{fig11}. The lines  represent  a large number of snapshots with  low scattering. There is a clear    anticorrelation of turbulence intensity and  Prandtl number $\Prr$. The smaller the ordinary Prandtl number the larger is the Mach number of the convection.

Figure \ref{fig10} demonstrates the influence of the rotation on the  horizontal turbulence intensity $\langle u_x^2\rangle$ of the thermal convection which simultaneously represents its mean-field pressure. The black solid line  of the top plot gives the averaged quantity of Fig. \ref{fig11}. This curve is anti-quenched by slow rotation and it is quenched for rapid rotation just as it was discussed below Eq. (\ref{ux30}) for driven turbulence within the quasilinear theory.  We take this result as a strong  motivation for further applications of the SOCA theory also for rotating MHD models. The bottom panel displays the same numerical models  with respect to  the difference  
$\langle u_x^2\rangle- \langle u_x^{(0)2}\rangle$ which  gives  the excess of  turbulent pressure with respect to the non-rotating convection.  Negative values describe the rotational quenching of the turbulence  intensity by the rotation while positive values describe a support of 
$\langle u_x^2\rangle$. The latter  happens for slow rotation (red and green lines, $\langle u_x^2\rangle>\langle u_x^{(0)2}\rangle$) while for rapid rotation with  $\langle u_x^2\rangle< \langle u_x^{(0)2}\rangle$  the opposite is true. 

The total pressure by rotating convection is the sum of the turbulence-originated pressure $\langle u_x^2\rangle$   and the centrifugal pressure    $\Om^{*2}R^2/2D^2$ by the global solid-body rotation (in code units).  Obviously, for slow rotation the pressure term  $\langle u_x^2\rangle$   does not `feel' the rotation but it is (slightly)  reduced if it is fast. In the latter case the effective pressure is smaller than $ \langle u_x^{(0)2}\rangle$ plus centrifugal term.

\subsection{Magnetoconvection}\label{magnetocon}
\begin{figure}
 \centering
\hbox{
\includegraphics[width=4.3cm]{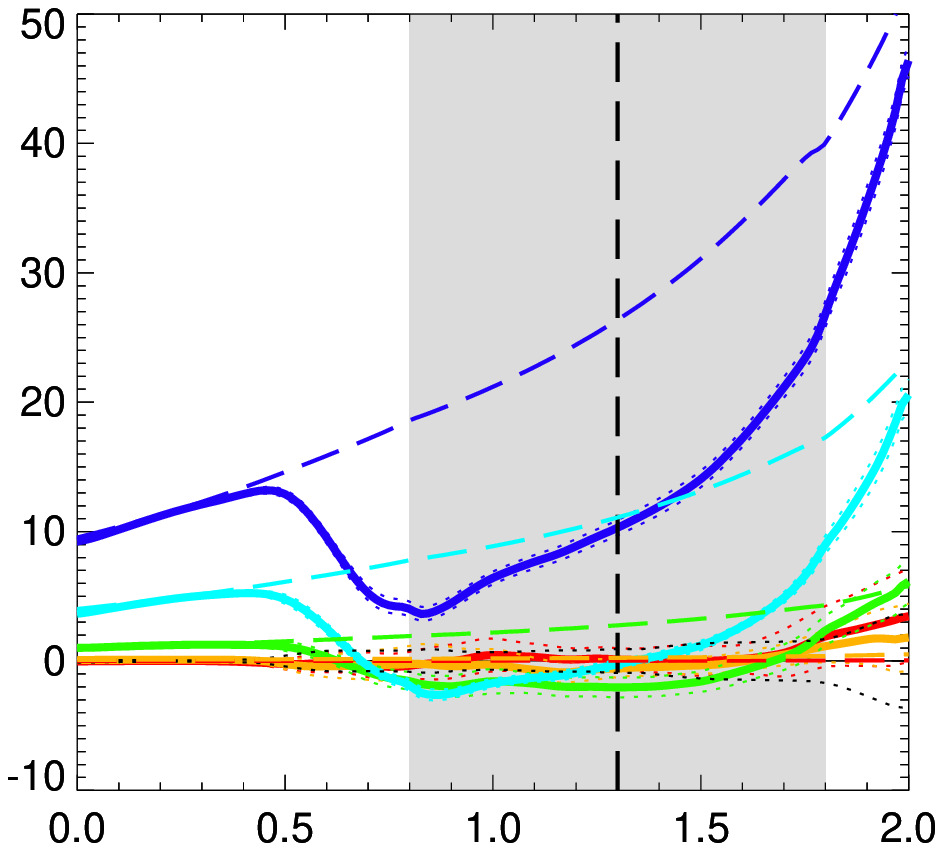}
\includegraphics[width=4.3cm]{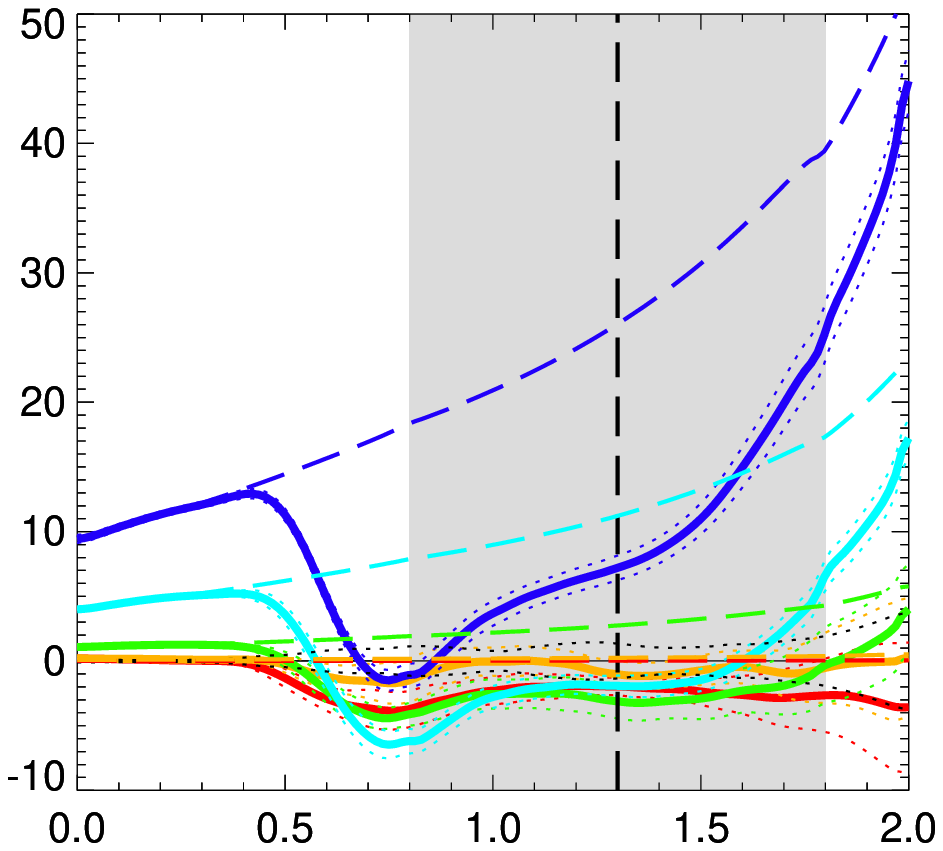}}
\caption{The pressure excess  $\delta P^{\rm tot}/\rho$  for vertical background field with $B^*=1$ (yellow), $B^*=3$ (red) and $B^*=10$ (green),  $B^*=20$ (light blue) and $B^*=30$ (dark blue). Negative values stand for $P_{\rm in}>P_{\rm ext}$  while positive values stand for  the standard relation  $P_{\rm in}<P_{\rm ext}$, see Eq.  (\ref{Dicke1}).    The dashed lines  give the pressure excess without turbulence in the same units as used in Fig. \ref{fig11}.   The vertical dashed line marks  the center of the unstable box at $z=1.3$ where the pressure excess is minimal for $B^*=10$.   $\Om^*=0$,  $\Pm=0.1$, $\Prr=0.1$ (left), $\Prr=0.05$ (right).}
\label{fig13}
\end{figure}

The turbulence intensity $\langle \vec{u}^2\rangle$  for $B^*=10^{-3}$ and $B^*=1$ has  been computed without rotation by \cite{RK12}   without  remarkable differences of the two models. The influence of vertical magnetic fields upon the pressure should thus be negligible for $B^*< 1$.  
 This is indeed shown by the red line of Fig. \ref{fig13}. Magnetic fields must exceed  
  $B^*\simeq 1$ in order to  influence the convective  pressure remarkably.
  However, fields with $B^*>1$ indeed   reduce the magnetic pressure   (\ref{Dicke1}) without turbulence. Figure \ref{fig13} gives the pressure excess  (\ref{ptot30})  without rotation  for vertical background fields with $B^*\leq 30$.  Negative values stand for negative pressure excesses, $P_{\rm in}>P_{\rm ext}$, which appear for both the given Prandtl numbers for moderate $B^*$ green and light blue lines). The Prandtl numbers are similar to those used by \cite{KBK12}.  
 
 In order to exclude boundary effects we focus  attention to the values around the central line at $z=1.3$ (vertical dashed lines). For comparison the   dashed lines  give the pressure excess {\em without turbulence} in the same units as used in Fig. \ref{fig11}. They represent the  solution of Dicke in Eq. (\ref{Dicke1}). Because of the density stratification the curves are {\em not} strictly horizontal\footnote{the normalized initial density is set to unity at the upper boundary of the unstable layer}. Values of the pressure excess  below the  dashed lines stand for magnetic pressure suppression ($\kappaP>0$) while  values above the dashed lines reflect  magnetic-induced pressure amplification ($\kappaP<0$). In all  cases plotted in Fig. \ref{fig13}  a distinct reduction of the large-scale magnetic pressure  by the turbulence within the unstable zone appears independent of the  Prandtl number.   For  strong fields with $B^*>20$, however,
  the  amplitudes are no longer large enough  to generate negative values of the pressure excesses. The figure demonstrates  the existence of a maximal magnetic field strength  of about $B^*=20$ for the negative-pressure effect. For stronger fields the turbulence-free magnetic   term in (\ref{Dicke1}) can never be over-compensated by the turbulence quenching.  With the limiting value $B^*=20$ the ratio of the magnetic energy to the kinetic energy is $\simeq 0.4$ (see Fig. \ref{fig11}). One finds that the turbulence can indeed reduce this value or can even change the sign  of $\delta P^{\rm tot}/\rho$.  After the second relation in (\ref{Dicke1}) also the temperature excess would then change its sign  and in the magnetized domain   becomes bright rather than  dark. 
  \begin{figure*}
 \centering
\mbox
{\includegraphics[width=6cm]{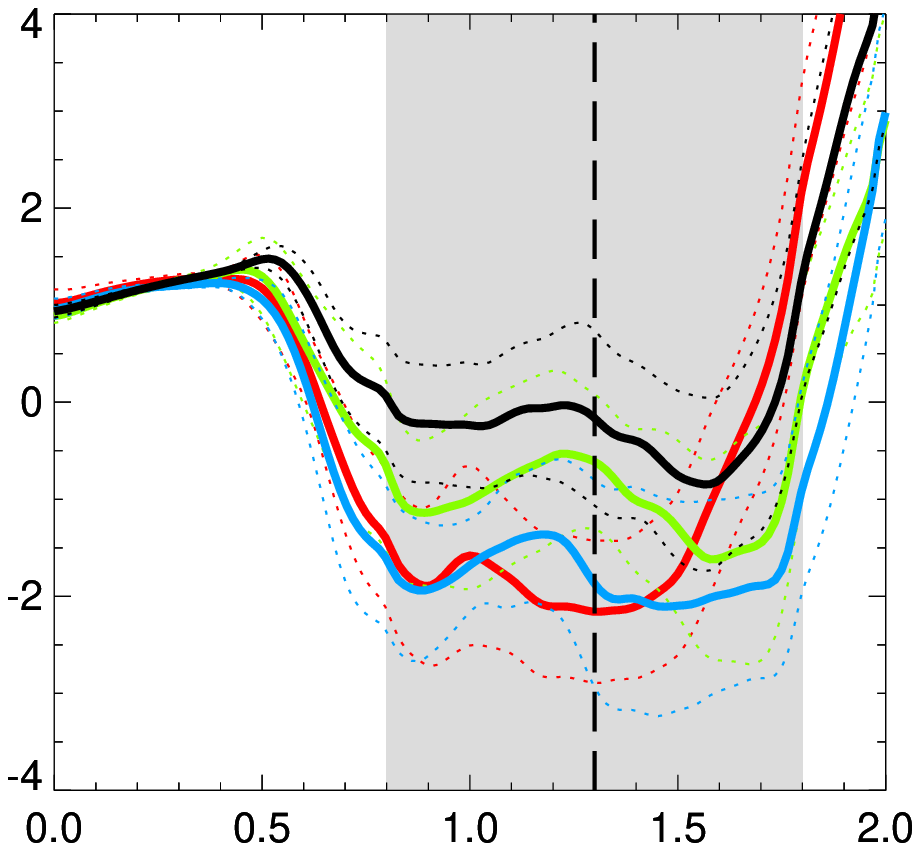}\hfill
\includegraphics[width=6cm]{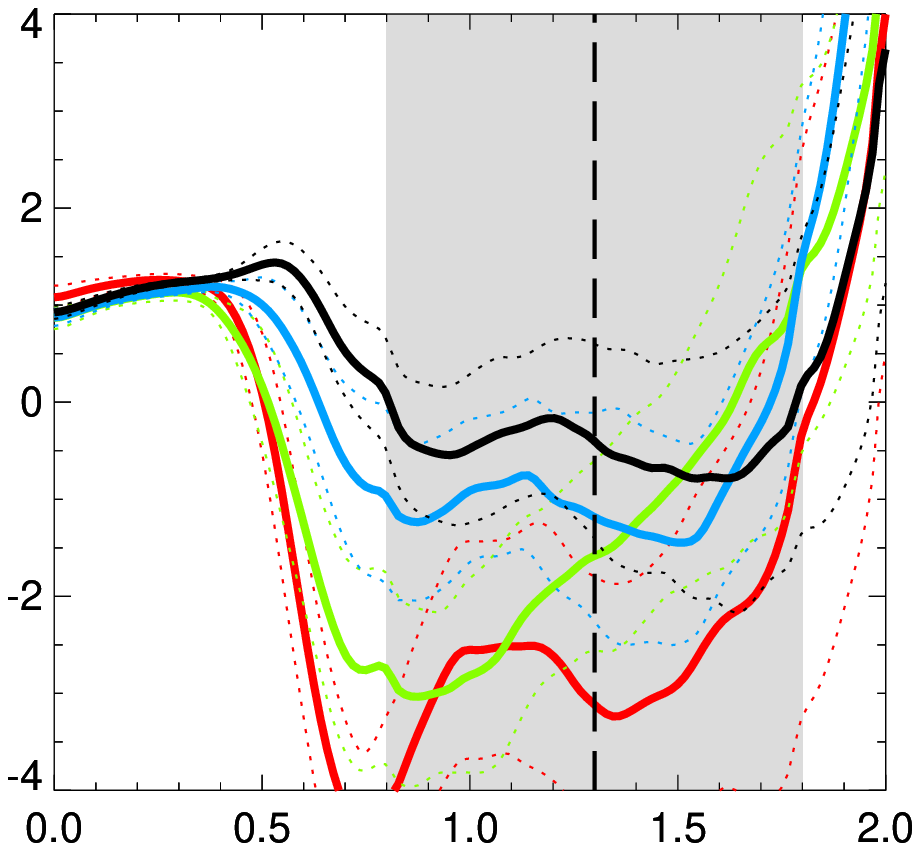}}
\caption{Rotating magnetoconvection with  $\Mm<1$ for $\theta=0^\circ$:  pressure excess $\delta P^{\rm tot}/\rho$  after Eq. (\ref{ptot30})  for $\Om^*=0$ (red), $\Om^*=1$ (green),   $\Om^*=3$ (blue)  and $\Om^*=10$ (black)  for vertical background field with $B^*=10$.  Negative values stand for magnetic  quenching while positive values stand for  anti-quenching. $\Pm=0.1$.  $\Prr=0.1$ (left),  $\Prr=0.05$ (right).}
\label{fig30}
\end{figure*}   

In the simulations by \cite{KBK12}  where for weak fields  the turbulence  always leads to negative magnetic-pressure effect the ordinary Prandtl number is fixed to   $\Prr= 1$.  In our models 
decreasing Prandtl numbers    imply decreasing magnetic resistivity $\eta$: the values in code units are $\eta=0.063$ for $\Prr=0.1$ and $\eta=0.044$ for $\Prr=0.05$. 
The locations of  the  lines, however,  do hardly depend on the value of the ordinary Prandtl  number.  In all cases  the fields with $B^*=10$ lead to a  significant  {\em amplification}  of the inner molecular pressure by about 7\%. We shall show, however,  that under the influence of a global rotation   these  models 
loose their negative performance.  For weaker magnetic fields   the effects are weaker. 
 
 \subsection{Rotating magnetoconvection}   

So far the modification of Eq. (\ref{Dicke1}) under the presence of rotating or magnetized turbulence has been considered. The natural next question is that after the structure of rotating {\em and} magnetized  turbulence. Equations (\ref{qqs})--(\ref{p2}) form  the  expressions for the numerical simulations. The combined  influences of magnetic field and rigid rotation with parallel $\vec\Om$,  $\vec B$ and $\n\rho$ are probed   for slow and  rapid rotation. 

After   Fig. \ref{fig13} the nonrotating models with    $B^*=10$ (green lines) provide the largest negative turbulent pressure excesses. We shall thus use this value  in order to study the rotational influence.  It is known  from Fig. \ref{fig10}  that  rotation  amplifies the horizontal turbulence intensity  hence  we expect the  lines for $\Om^*\neq 0$ to be shifted upwards in comparison with the red lines for $\Om^*=0$.  This  is indeed the result of the simulations shown in the plots of Fig.  \ref{fig30} for two values of the ordinary Prandtl number.  The rotation  over-compensates the negative-pressure effect which  for $\Om^*\gsim 10$ starts to disappear. 
Hence,  for $\rm Mm=1$ it is in the center of the box    $\delta P^{\rm tot}\simeq 0$. Hence, if the magnetic background field is parallel to the rotation axis the interaction of magnetic  field and  rotation keeps the resulting  differences of  gas  pressure    for magnetic and non-magnetic convection as rather  small.

\begin{figure}
\vbox{
\hbox{\includegraphics[width=4.3cm]{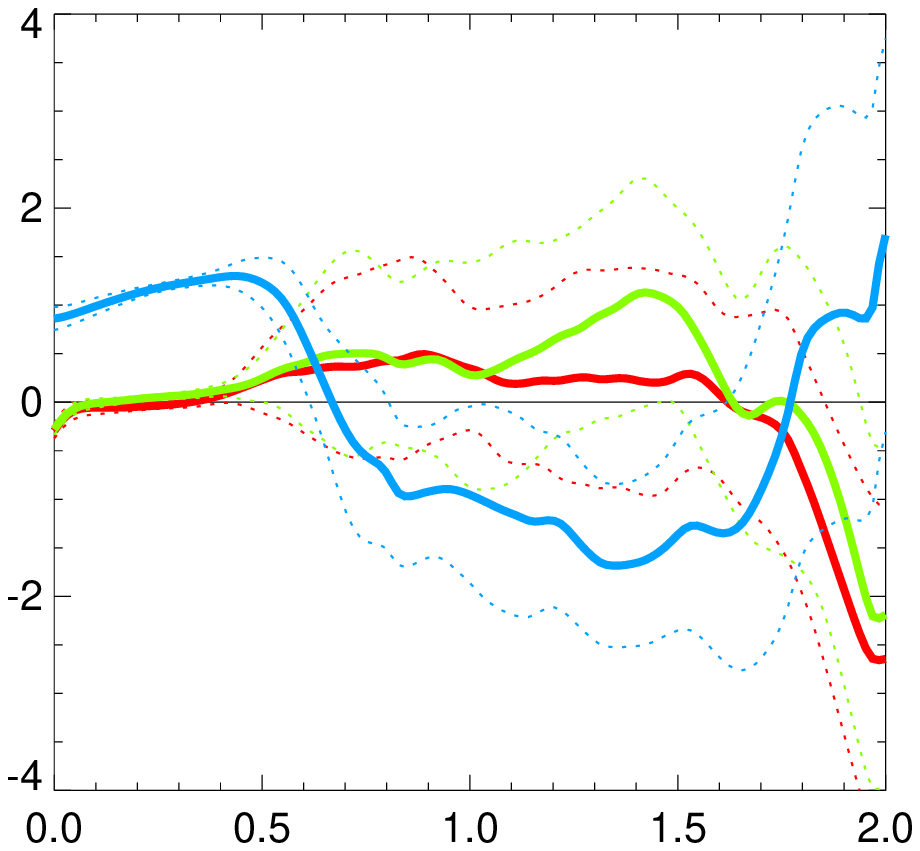}
\includegraphics[width=4.3cm]{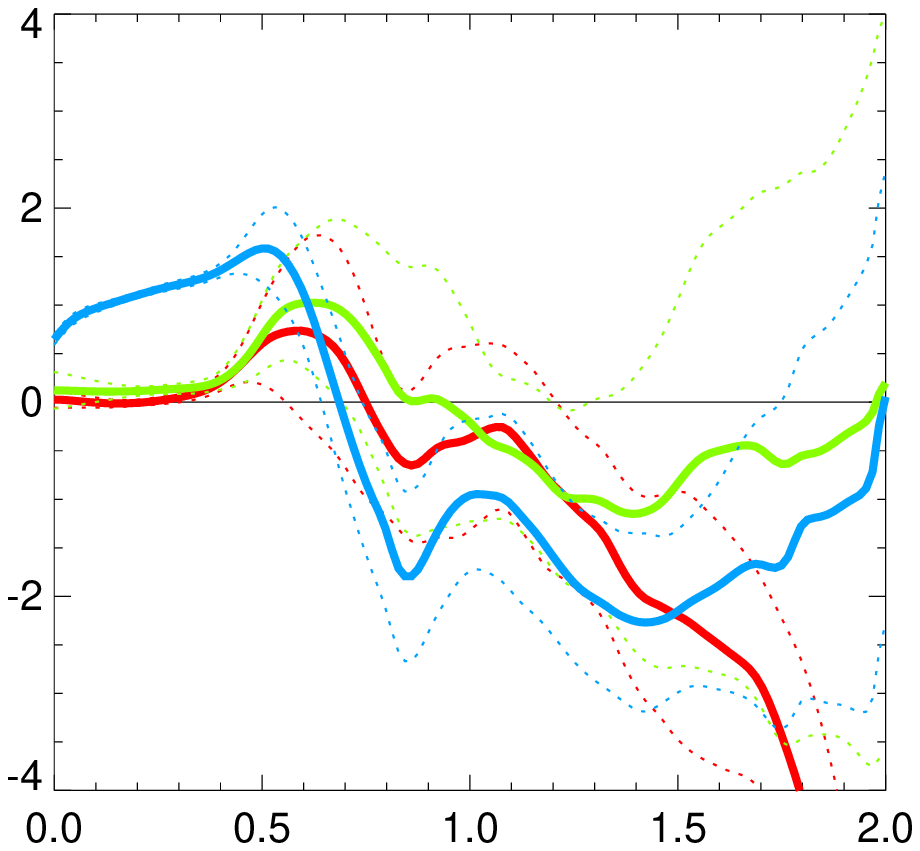}}
\hbox{
\includegraphics[width=4.3cm]{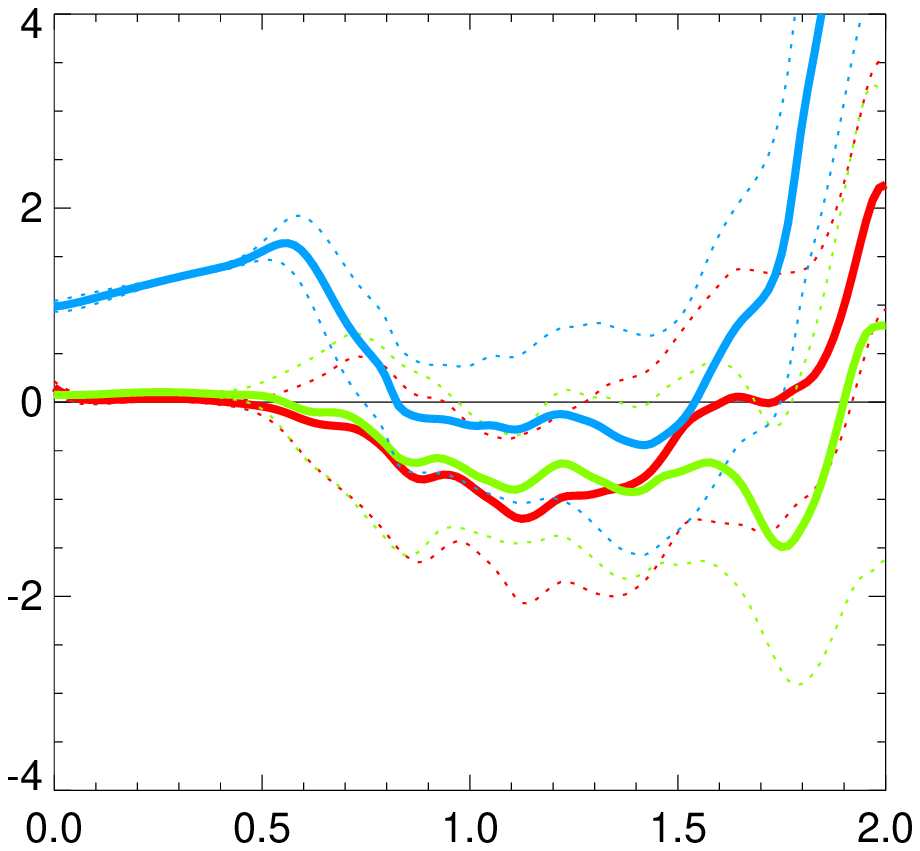}
\includegraphics[width=4.3cm]{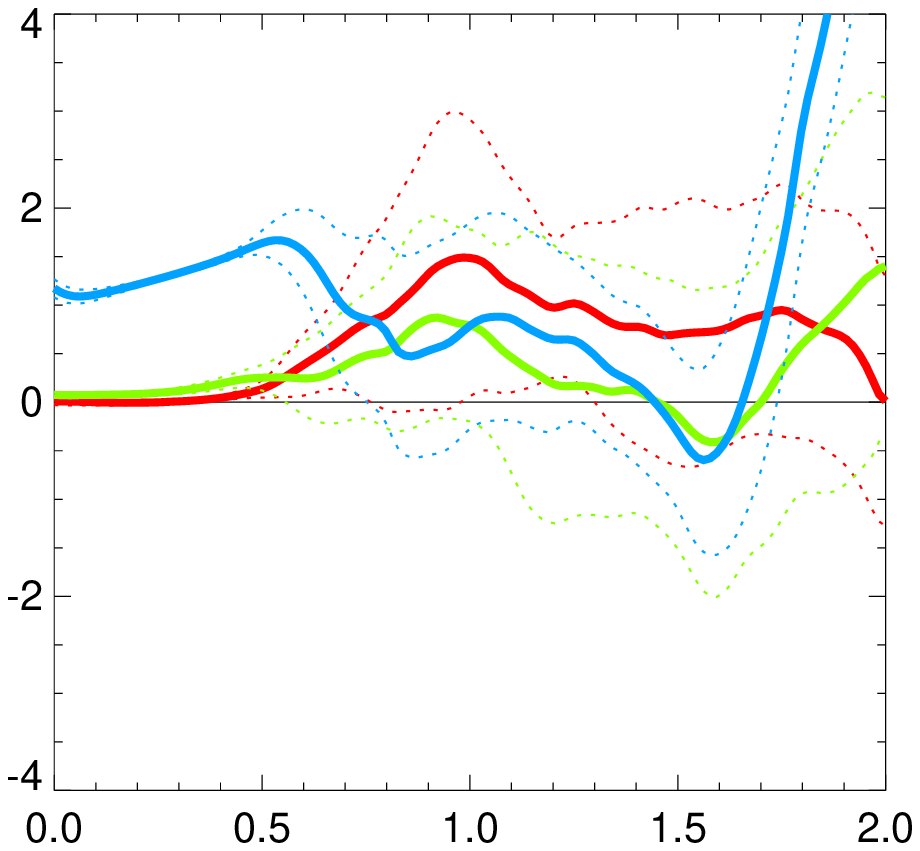}}
}
\caption{The total pressure excess   $\delta P^{\rm tot}/\rho$  for  vertical background field $B^*=1$ (red), $B^*=3$ (green), $B^*=10$ (blue) for oblique rotation with  $\theta=30^\circ$ (left) and $\theta=45^\circ$ (right). Top: $\Om^*=1$, bottom:  $\Om^*=10$. 
$\Pm=\Prr=0.1$.}
\label{fig21}
\end{figure}

The calculations are  more complicated for rotating magnetoconvection if a finite angle exists between the magnetic field and the rotation axis. One should believe  that the effective rotation rate for the convective  box with vertical magnetic field is reduced by the factor $\cos{\theta}$ but this is only one side of the medal. On the other hand, it is also true that the magnetized oblique rotator generates a turbulence field which is anisotropic also in the horizontal ($x,y$) plane.   The numerical simulations provide a rather clear picture. Figure \ref{fig21} shows the results for  inclination angles $\theta=30^\circ$ (left panel) and $\theta=45^\circ$ (right panel).  In both cases for rapid rotation (bottom plots) the lines for the three magnetic field amplitudes proceed to positive  values. The negative-pressure effect also disappears for rapid rotation, therefore,  when the magnetic axis and the rotation axis are not parallel.
\subsection{Rapid rotation, $\Mm>1$}
\begin{figure*}
 \centering
 \mbox{
\includegraphics[width=6cm]{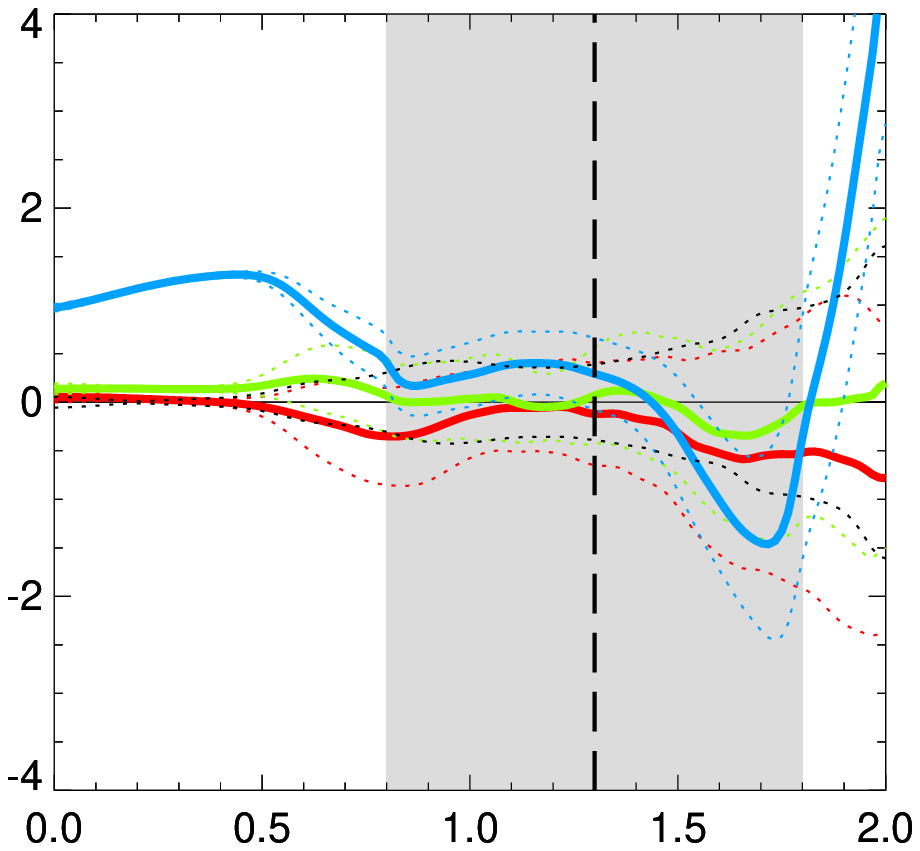}\hfill
\includegraphics[width=6cm]{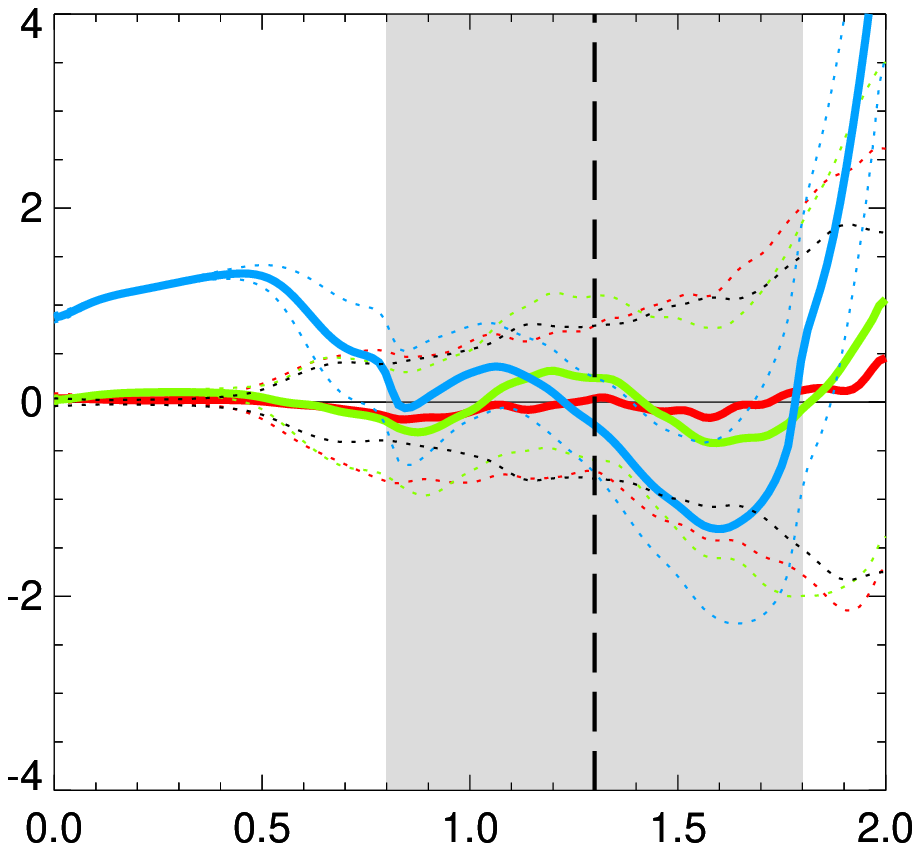}}
\caption{Rapidly-rotating magnetoconvection with $\Mm>1$ for $\theta=0^\circ$ (left) and $\theta=30^\circ$ (right). Pressure excess $\delta P^{\rm tot}/\rho$  after Eq. (\ref{ptot30})  for $\Om^*=30$.  $B^*=1$ (red),   $B^*=3$ (green)  and $B^*=10$ (blue). The pressure excess almost completely disappears. $\Pm=0.1$, $\Prr=0.1$.}
\label{fig32}
\end{figure*}

It remains to study the pressure differences in rapidly rotating convection  with 
magnetic Mach numbers exceeding unity. Our model works with $\Om^*=30$ with $B^*=1$, $B^*=3$ and $B^*=10$ so that the magnetic Mach number is $\Mm>1$ reaching values up to 30. Figure \ref{fig32} demonstrates how the pressure excess goes to zero for increasing magnetic Mach number both for $\theta=0^\circ$ and for $\theta=30^\circ$.  In these cases, therefore,  Eq. (\ref{Dicke2}) simplifies to $P_{\rm in}\simeq P_{\rm ext}$ despite the existence of large-scale  magnetic fields. The explanation of the darkness of  sunspots by the mean Lorentz force inside the spot domain, if real, could    work for rapid rotation. Also the opposite assumption of negative pressure excess inside the sunspot (and possible resulting instabilities) does not hold for rapid rotation. 
For fast rotating turbulence  the sum of Reynolds stress and Maxwell stress does not depend on the strength of the (moderate) magnetic field. The total pressure does thus not depend on the magnetic field (if $B^*\leq 10$), it always equals the pressure without magnetic field. Weak and moderate fields do neither reduce nor enhance the turbulence pressure, with respect to the pressure the magnetic field is hidden by the turbulence. 

If  starspots of similar structure exist for both slow rotation and rapid rotation then the turbulence-originated negative magnetic-pressure excess can not be essential for the spot-formation. 
\section{Conclusions}
The total pressure  in  a rotating  and/or  magnetized  convection is considered where mostly the  rotation axis, the magnetic field and the density gradient are parallel. We start to discuss the anisotropies originated by rotation or by magnetic field for driven turbulence in a quasilinear approach. Often the results of such analytic correlation approximation are confirmed by direct numerical simulations but we also have to look for possible differences. Note that the quasilinear approach  only deals with driven turbulence.

If the driven turbulence is isotropic without rotation it is anisotropic with rotation in the sense that the  turbulence intensity transverse to the rotation axis exceeds the turbulence intensity along the rotation axis. We find the opposite anisotropy  for driven turbulence subject to a uniform axial magnetic field.  It is thus not surprising that the turbulent pressure (by the Reynolds stress) for rotation exceeds the turbulent pressure without rotation and again the opposite is true for the influence of axial magnetic fields, see Eq. (\ref{ux30}). 

However, the total pressure in turbulence under the influence of magnetic fields is formed not only by the Reynolds stress but also by the Maxwell stress where the latter is combined by  a large-scale and a small-scale   part. All together form the total pressure excess $\delta P^{\rm tot}= P^{\rm tot}(B)-P^{\rm tot}(0)$ which in equilibrium equals the gas pressure difference without and with magnetic field. Without turbulence   $\delta P^{\rm tot}$ is positive (almost) by definition, see Eq. (\ref{Dicke1}). It is confirmed that for turbulence under the influence of weak magnetic fields  $\delta P^{\rm tot}$ can be negative which phenomenon is  called the negative magnetic-pressure effect \citep{BK10,BK11,BK12}.

One finds from the SOCA approximation for driven turbulence that for weak fields this effect should not exist for large $\Pm$. Moreover, for $\Pm<8$ upper bounds of the magnetic Reynolds number of the turbulence exist to allow negative $\delta P^{\rm tot}$. For higher magnetic Reynolds numbers, however,  it is always  $\delta P^{\rm tot}>0$. The negative magnetic-pressure effect can   thus only exist for rather low values of the molecular electric conductivity.  For the  high-conductivity limit  with  $\mu_0\sigma \ell^2_{\rm corr}\gg\tau_{\rm corr}$  the  turbulence-originated pressure {\em enhances} the large-scale magnetic pressure $B_0^2/2\mu_0$ rather than to reduce it.
For decreasing     
Prandtl number $\Prr$ the magnetic-free turbulence intensity slightly  grows (Fig. \ref{fig11}) while simultaneously  for the MHD models the molecular   $\eta$ sinks. 


For the model of the lowest curve in Fig. \ref{fig11}  we have computed the correlation time by means of its autocorrelation function. The result is $\hat \tau\simeq 0.1$ in code units. Hence $\etaT\simeq \langle u_x^2\rangle {\hat  \tau}  \simeq 2.5$ taken in the middle of the convection box. For the  correlation length it results $\ell_{\rm corr}\simeq \sqrt{\langle u_x^2\rangle} \hat\tau\simeq 0.5$. In the average, the box contains two eddies in the vertical dimension. With $\eta=0.063$  taken from Section \ref{magnetocon} one finds $\etaT/\eta\simeq 40$ as a   proxy of  the magnetic Reynolds number $\Rm'$ of the fluctuations.

For vertical fields it is even possible to  derive the eddy diffusivity  $\etaT$  directly from the simulations. We have shown earlier that convection subject to uniform magnetic  fields provides  a finite cross correlation $\langle \vec{u}\cdot \vec{b}\rangle$ proportional to the scalar product $ {\vec B}\cdot\n \rho$ \citep{RK12}. It vanishes for homogeneous turbulence and for fields perpendicular to the density stratification.
The quasilinear approximation {and} numerical simulations lead to  
\beg
	\langle u_z b_z\rangle = -\etaT \frac{ B_0}{H_\rho} ,
	\label{cross}
\ende
where $H_\rho$ as the density scale height is here about 1.4 in code units for all  models. 
We have computed the correlation  (\ref{cross})  in the middle plane of the convective box averaging over the horizontal plane and time.  The resulting eddy diffusivity  values for $B^*=3$ and $B^*=10$ are almost  identical for one and the same model.  We find $\etaT/\eta\simeq 30$ for both $ \Prr=0.1$ and $\Prr=0.05$ close to the above given approximative result. 
For $\Prr<0.05$ the reliability of the simulations with our code was only restricted.

It remains to report the influence of rapid rotation to the low-conductivity case  where the negative magnetic-pressure effect for moderate fields  always exists.  Figure \ref{fig30} demonstrates that  it  is  erased by the basic rotation.   For  $\Mm> 1$
 the pressure differences between magnetic and non-magnetic fluids  are completely planished (Fig. \ref{fig32}). For  large rotation rates  it is  always $P^{\rm tot}(B)\simeq P^{\rm tot}(0)$ so that the influence of the (weak) magnetic field disappears. For all fields the sum of turbulent pressure  and magnetic pressure  (the large scale contribution included) remains constant and does not depend on the magnetic field amplitude. 
 
 It is often argued that large-scale magnetic fields and turbulence  are supporting cold molecular clouds 
 against self-gravity and the external pressure.  The star formation rate seems to be lowered  by the magnetic pressure so that only a few percent of the mass of the molecular cloud reaches a stellar configuration. The observed large-scale magnetic fields are of order 30 $\mu$G on scales of 0.1-10 pc \citep{C99}. On the other hand, the angular momentum transport  by the Lorentz force \citep{D81} and/or magnetic quenching of the turbulence \citep{ZV18} may increase the star formation rate. Our result that for large magnetic Mach numbers the total pressure (combined by turbulence plus small-scale magnetic fields plus large-scale magnetic field) always equals its non-magnetic value  excludes a basic magnetic influence on the star formation rate as long as the magnetic field is weak. It is not finally clear, however, if it is allowed to transmit  the properties  of  rotating convection to other forms of turbulence.
 
 Another open  question is whether not only the magnetic-originated diagonal elements of the tensor (\ref{3}) but also its off-diagonal elements vanish for rapid rotation. Of particular interest are the terms with one of the indices representing  $\phi$ (in spherical coordinates) which describe  the angular momentum transport by magnetic fields and turbulence. One can show that the angular momentum transport by rotating density-stratified turbulence vanishes for $\Om^*\to \infty$ even under the presence of an azimuthal magnetic field  \citep{KR94} but
 the consequences of fast-rotating turbulence for the magnetic braking by the background Lorentz force are still unknown.

\bibliography{pressure}

\end{document}